\documentstyle[12pt]{article}
\begin{document}
  
\centerline{\normalsize\bf NEUTRINO MASS IN PHYSICS AND ASTROPHYSICS}
\baselineskip=16pt

\centerline{\footnotesize LINCOLN WOLFENSTEIN}
\baselineskip=13pt
\centerline{\footnotesize\it Department of Physics, Carnegie Mellon
University,}
\baselineskip=12pt
\centerline{\footnotesize\it Pittsburgh, PA~~15213, USA}
\centerline{\footnotesize E-mail: wolfenstein@cmphys.phys.cmu.edu}
\vspace*{0.3cm}
\vspace*{0.9cm}
  \begin{abstract}{The symmetry between quarks and leptons suggests that
neutrinos should have mass.  As embodied in the grand unified theory
SO(10) this yields masses that can only be detected by neutrino oscillations. 
Such oscillations could be very important for supernova physics.  Present
observations of solar neutrinos when combined with standard solar model
calculations imply particular parameters for neutrino masses and mixings.  If the
solar model is somewhat relaxed quite different possibilities emerge, which
yield very different predictions for future experiments.}
\end{abstract}
   
  \normalsize\baselineskip=15pt
  \setcounter{footnote}{0}
  \renewcommand{\thefootnote}{\alph{footnote}}
  \section{Introduction}
 
It is sometimes difficult to explain to those outside our field why neutrinos
are so important.  They interact so weakly that they do not bind to matter and
are extremely difficult to detect.  Yet it is just because they interact only
weakly that they are so important.  On the one hand in particle physics they
allow us to probe weak interactions and perhaps interactions beyond the
standard model.  On the other hand in astrophysics neutrinos can emerge from
dense regions from which there is no other direct information.  In this talk I
focus on neutrino mass which will play a large role in our discussions here.

\vspace{0.25in}

\noindent {\bf 2.~~Neutrino Mass and Particle Physics}

\vspace{0.16in}

When parity violation was discovered in 1957 it was suggested that the neutrino
was a Weyl particle, having only two states $\nu_L$ and $\overline{\nu}_R$. 
Assuming a distinction between particle and anti-particle it then had to be
massless.  However in the standard model all particles start out as massless
Weyl particles, left-handed doublets and right-handed singlets.  In fact
$\nu_{eL}$ and $e_L$ form such a doublet.  When the SU(2) symmetry is broken
the left-handed particles are connected to the right-handed particles (via the
Yukawa interaction) to form four-component Dirac particles.  Thus the question
becomes whether one should introduce or leave out the right-handed neutrino in
the cast of characters.  It is customary to leave it out, but this is perfectly
arbitrary as far as the standard model goes.

Is there some reason to introduce a right-handed neutrino?  I think there is. 
It has to do with the approximate symmetry one observes between quarks and
leptons.  They come in three generations.  The charged leptons $(e,\ \mu,\
\tau)$ have a mass hierarchy resembling the quarks $(d,\ s,\ b)$.  The weak
interactions of quarks and leptons are the same.  Thus a very attractive idea
is that this is a broken symmetry which becomes exact at some high mass scale. 
This symmetry was labeled the SU(4) of color by Pati and Salam
~\cite{Pati/Salam} with the lepton being the fourth color.  When the symmetry
holds there must be right-handed neutrinos.

The grand unified theory (GUT) SO(10) has this symmetry.  Quarks and leptons
are different components of the (16) representation.  As part of the
symmetry breaking Gell Mann, Ramond, and Slansky ~\cite{Mann/Ramond/Slansky}
suggested that $\nu_R$ obtains a large Majorana mass $M$.  The normal mass
term that mixes $\nu_R$ with $\nu_L$ (with a magnitude $m_D$ comparable to the
masses of up-type quarks) now causes a small mixing of the heavy $\nu_R$ into
$\nu_L$ giving the light neutrino a small Majorana mass

\begin{equation}
m_{\nu} = m^2_D/M
\end{equation}

\noindent This is the famous see-saw formula.  From this point of view the
magnitude of the neutrino mass is a clue to the scale at which the quark-lepton
symmetry is broken.

Unfortunately it is not possible to derive quantitative results on neutrino
masses and mixing from SO(10) without making detailed assumptions about the
forms of mass matrices.  This has been done in many papers~\cite{Babu/Mohapatra}
and most, but not all, give the general qualitative results:

(1) There exists a mass hierarchy $m\left(\nu_3\right) \gg
m\left(\nu_2\right)\gg m\left(\nu_1\right)$  where $\nu_3,\ \nu_2,\
\nu_1$ have as their major component $\nu_{\tau},\ \nu_{\mu},\ \nu_e$.

(2) There exists neutrino mixing analogous to quark mixing so that $\nu_e =
U_{e1}\ \nu_1 + U_{e2}\nu_2 + U_{e3} \nu_3$ with $U_{e1} \sim 1,\ U_{e2}$ small
but not extremely small, and
$U_{e3} < U_{e2}$.

(3) The scale $M$ is greater than $10^{10}$ Gev so that all neutrino masses are
small and can only be explored via neutrino oscillations.

While it is natural in SO(10) to obtain neutrino mass, it is possible by adding
a second set of right-handed neutrinos to obtain zero masses
~\cite{Wyler/Wolfenstein}.

\vspace{0.25in}

\noindent {\bf 3.~~Supernovae and Neutrino Mass}

\vspace{0.16in}

When massive stars complete their nuclear burning they suffer a catastrophic
collapse in which an enormous amount of energy is generated.  Calculations have
shown that nearly all this energy is emitted in the form of neutrinos because
they can escape most easily.  In one of the great events of modern science
about 18 of these neutrinos were observed in the 1MB and Kamiokande detectors
from supernova 1987a.  This observation wonderfully confirmed our picture of
supernovas as well as constrained hypothetical new particles.  It still amazes
me that 150,000 years ago these neutrinos set out timed to arrive just a few
years after these detectors were set up.

The supernova watch remains one of the major tasks of neutrino telescopes.  A
supernova in our own galaxy can be studied even if its light is obscured.  All
detectors capable of measuring these neutrinos should be ready as much of the
time as possible and have accurate absolute timing.

The surface of the collapsed star from which the neutrinos emerge is called the
neutrinosphere.  Neutrino oscillations that may occur after the neutrinos
emerge can be very significant.  Although all three types of neutrinos emerge,
$\nu_{\mu}$ and $\nu_{\tau}$ neutrinos have higher energies than $\nu_e$
because they come from deeper within the neutrinosphere since their
cross-sections are lower.  Thus neutrino oscillations from $\nu_{\mu}$ or
$\nu_{\tau}$ to $\nu_e$ have the effect of increasing the $\nu_e$ energy.

As far as SN1987a is concerned all or nearly all of the neutrinos observed are
believed to be $\overline{\nu}_e$ which have by far the largest cross-section in
the detectors.  If there were a large mixing of $\overline{\nu}_{\mu}$ to
$\overline{\nu}_e$ the detected energies would have been greater.  This has
been used by some authors to rule out large mixing
~\cite{Smirnov/Spergel/Bahcall} (such as in the vacuum oscillation solution for
solar neutrinos), but the conclusion is statistically limited.

Assuming $\nu_e$ is the lighest neutrino there is much interest in the
possibility of MSW oscillations transforming $\nu_{\mu}$ or $\nu_{\tau}$ to
$\nu_e$ as the neutrinos pass through a great range of density from the
neutrinosphere to the expanding surface of the star ~\cite{Schramm/Walker}.  The
higher energy of the resulting $\nu_e$ could have an important effect on explosion
calculations.  Also the higher energy $\nu_e$ could change neutrons to protons
creating a problem for explosive nucleosynthesis ~\cite{Fuller/Qian}.

\vspace{0.25in}

\noindent{\bf 4.~~Solar Neutrino Problem}

\vspace{0.16in}

I will briefly review the standard discussion of solar neutrinos that will
occupy much of this meeting.  The source of the sun's energy was identified in
the 1930's as nuclear reactions in the hot core of the sun.  Neutrinos that can
penetrate from the center to the surface of the sun provide the one direct way
of detecting these reactions.  The detection of solar neutrinos has provided a
wonderful confirmation of this general picture.

The standard solar model (SSM) calculations indicate that for a star with the
mass of the sun nearly all the energy comes from the {\it PP} cycles.  These lead
to three important sets of neutrinos:

(1) $pp$ neutrinos with a  continuous spectrum up to 420 kev.  These originate
from the primary weak interaction

$$p + p \rightarrow d + e^+ + \nu_e$$

\noindent The flux of these neutrinos depends very little on  details of the SSM
because it is highly constrained by the observed solar energy production.

(2) $^7Be$ neutrinos with a line spectrum mainly at 860 kev arising from

$$e^- + ^7Be \rightarrow \nu_e + ^7Li$$

\noindent The flux of these neutrinos, an order of magnitude lower than the
pp, can change by the order of 20\% between different versions of the SSM. 
It is approximately sensitive to $T_c^{10}$, where $T_c$ is the central
temperature~\cite{Bahcall/Ulmer}.

(3) $^8B$ neutrinos with a continuous spectrum up to 15 Mev arising from 

$$^8B \rightarrow ^8Be + e^+ + \nu_e$$

\noindent This flux, which is four orders of magnitude lower than the $pp$
flux, is the most uncertain because it depends on the rate of formation of $^8B$
from $p + ^7Be$.  It is approximately sensitive to $T_c^{24}$ and the
uncertainty may be as large as 50\%.

There are three types of experiments that have now detected solar neutrinos:

(1) Water Cerenkov detectors detect neutrinos via neutrino-electron
scattering.  The threshold for the Kamiokande experiments is 7.5 Mev which
means they are sensitive only to $^8B$ neutrinos.  The observation of a signal
by this detector showed that indeed the rare {\it PP} III branch occurs.  The
detected rate is about 50\% of the SSM.~\cite{Bahcall/Pinsonneault}  This by
itself does not represent an extreme problem given the SSM uncertainties.

(2) Chlorine radiochemical experiment.  This experiment is calculated to be
primarily sensitive to $^8B$ neutrinos (6.2 SNU) but its threshold also allows
a significant signal (1.2 SNU) from $^7Be$ as well as a small signal (0.6 SNU)
from {\it pep} and CNO neutrinos.  The observed rate (2.55 $\pm$ 0.26) SNU is
about $\frac{1}{3}$ of the SSM.  One can use the Kamiokande experiment to predict
that $^8B$ neutrinos should give a signal of at least 2.6 SNU and this leaves
no room for the $^7Be$ neutrino signal.  It is this comparison that first clearly
seemed to indicate a solar neutrino problem.

(3) Gallium radiochemical experiment.  The great importance here is
that 74 SNU of the expected signal (131 SNU) comes from {\it pp} (plus {\it pep}
neutrinos) for which the rate is highly constrained by the observed luminosity. 
The GALLEX observation of $77 \pm 10$ SNU as well as the SAGE result ($72 \pm
13$ SNU) is consistent with the expected {\it pp} flux but leaves no room for the
$^7Be$ signal expected to be 36 SNU.  This then can be considered a second solar
neutrino problem.~\cite{Bahcall}

A standard solution is the assumption of the MSW neutrino oscillations with
parameters that greatly suppress the $^7Be$ flux and also suppress the $^8B$
flux.  A preferred solution~\cite{Hata/Langacker} has $\Delta m^2 \approx
6.10^{-6}$ ev$^2$ and
$\sin^2 2 \theta \approx 10^{-2}$.  Within the theoretical framework discussed
above there are two scenarios:

Scenario A.  The scale M is of order $10^{12}$ GeV.  This
occurs~\cite{Deshpande/Keith/Pal} in non-SUSY models where it is suggested that
SO(10) breaks in two steps to best fit the data with the breaking of quark-lepton
symmetry occurring at $10^{12}$ GeV with the GUT scale of order $10^{16}$ Gev. 
The MSW solution then corresponds to $\nu_e - \nu_{\mu}$ oscillations with
$m(\nu_{\mu}) \approx 2.5
\cdot 10^{-3}$ ev.  One then expects $m(\nu_{\tau})$ above $10^{-2}$ ev.  A
value of $m(\nu_{\tau}) \sim 10^{-1}$ ev with large $\nu_{\mu} - \nu_{\tau}$
mixing could then explain the atmospheric data.  Alternatively a value of
$m(\nu_{\tau})$ of a few ev could serve as a component of dark matter.

Scenario B.  The SUSY-GUT model in which SO(10) breaks to the standard model in
a single step at the GUT scale of $10^{16}$ Gev.  It is then natural (but not
absolutely necessary)~\cite{Lee/Mohapatra} that M be of order $10^{16}$ Gev.  The
MSW solution could then be $\nu_e - \nu_{\tau}$ oscillations.  In this case
$m(\nu_{\mu})$ would be expected of order $10^{-4}$ ev or less and $\nu_e -
\nu_{\mu}$ oscillations (either vacuum or MSW) could also affect the solar
neutrinos.  In this scenario the only way to detect oscillations is the study
of solar neutrinos.

\vspace{0.25in}

\noindent{\bf 5.~~Determining the Solar Neutrino Fluxes}

\vspace{0.16in}

It is of interest to ask what we really know about the neutrinos arriving from
the sun on the basis of experiment alone relaxing the constraints from SSM
calculations.  This is particularly important in considering what can be
learned from future experiments.

We assume that the standard well-known nuclear reactions are the source of the
solar energy.  We also assume that the solar luminosity is approximately
constant over the time required for this energy to appear at the surface so
that the observed luminosity determines the total energy produced.  Finally we
allow for the possibility of neutrino oscillations.  With these assumptions
Bahcall, Fukugita, and Krastev~\cite{Bahcall/Fukugita/Krastev} recently showed
that it is possible that nearly all the energy originates from the CNO cycle and
that the gallium experiments detect neutrinos from $^{13}N$ and $^{15}O$ rather
than pp.  This requires a wild departure from the SSM.  I have looked at a very
interesting but much less extreme possibility~\cite{Yu/Smirnov} in which most or
practically all of the $\nu_e$ detected come from $^7Be$.  This is exactly the
opposite of the standard conclusion discussed in the previous section.

The first point to note is that the Kamiokande experiment is sensitive to
$\nu_{\mu}$ and $\nu_{\tau}$ as well as $\nu_e$ but with a cross-section about
six times lower.  We consider an initial $^8B$ flux 2 (to 3) times the SSM with
oscillations converting 90 (to 100\%) of $\nu_e$ to $\nu_{\tau}$.  In this case
most of the Kamiokande events are due to $\nu_{\tau}-e$ scattering.  The
$\nu_e$ flux arriving from $^8B$ is then less than 0.2 times the SSM so that
most (or practically all) the $^{37}Cl$ signal must come from $^7Be$.  (We
assume the CNO contribution is small as in the SSM.)  This requires that the flux
of
$\nu_e$ from $^7Be$ arriving is 1 (to 2) times the SSM.  Then it follows that
half (or practically all) of the signal in the gallium detector is due to
$^7Be\ \nu_e$ so that the signal due to {\it pp} $\nu_e$ is half (or much less)
of the SSM.

Qualitatively we can understand this in terms of the Scenario B discussed in
the last section.  The $\nu_e - \nu_{\tau}$ MSW oscillations suppress the $^8B$
neutrinos with $m(\nu_{\tau}) \sim 10^{-2}$ ev.  Then the $\nu_e - \nu_{\mu}$
MSW oscillations suppress the {\it pp} neutrinos with $m(\nu_{\mu}) \sim 10^{-4}$
ev.  Alternatively $\nu_e - \nu_{\mu}$ vacuum oscillations suppress the pp
neutrinos with $m(\nu_{\mu}) \sim 10^{-6}$ ev.  In these scenarios the pp
neutrinos are suppressed considerably more than the $^7Be$ neutrinos just the
opposite of the usual picture.

While this picture does not require an extreme departure from the SSM we have
no reason to believe it is true.  What is important is that it leads to very
different predictions for SNO and Borexino.  For SNO the ratio of neutral
current to charged current is greater than 10 to 1 in contrast to the usual
picture where it might be 2 to 1.  For Borexino, which looks for $^7Be$
neutrinos, the signal is equal (or twice) the SSM in contrast to the usual
picture where it is much less.

\vspace{.25in}

\noindent{\bf 6.~~Conclusion}

\vspace{0.16in}

The observations of neutrinos from the sun and from SN1987a are two of the
great scientific events of recent times.  They are the pioneering efforts
in neutrino astronomy.  The possibility of small neutrino masses leads to an
exciting interplay between particle physics and astrophysics.  Future
generations of experiments with higher statistics and more detailed
measurements are needed to restrict the theoretical possibilities in particle
physics, as well as solar and supernova physics.

This work was supported in part by the U.S. Department of Energy, Contract No.
DE-FG02-91ER40682.

\end{document}